\documentclass[12pt]{amsart}
\usepackage{amsaddr}
\usepackage[pdftex]{geometry}
\usepackage[italian]{babel}

\usepackage{color}
\usepackage{hyperref}
\usepackage{graphicx}
\usepackage{amssymb}
\usepackage{xskak}
\usepackage{chessboard}

\begin{document}

\title{Automatic verification and interactive theorem proving}

\author{Andrea Asperti}
\address{Dipartimento di Informatica: Scienza e Ingegneria - DISI\\
  Universit\`a di Bologna}
\email{andrea.asperti@unibo.it}

\thanks{To appear in ``Direzioni della ricerca logica in Italia, vol 2. H.Hosni, G.Lolli, C.Toffalori eds.}
\date{}

\maketitle

\begin{abstract}
  Automatic verification deals with the validation by means of
  computers of correctness certificates. The related tools, usually
  called {\em proof assistant} or {\em interactive provers}, provide
  an interactive environment for the creation of formal certificates
  whose correctness can be assessed in a purely automatic way.
  Such \mbox{systems} have applications both in mathematics, where
  certificates are proofs of theorems, and in computer science,
  where certificates testify the correctness of a given software
  with respect to its specification.
\end{abstract}

\section{Introduzione}
    Esiste un interessante sito web\footnote{\url{https://www.win.tue.nl/~gwoegi/P-versus-NP.htm}} curato da G.J.Woeginger che raccoglie oltre un centinaio di contributi
    di autori che asseriscono di risolvere {\em in un senso o nell'altro} (!) il
    problema dell'uguaglianza tra le classi di complessit\`a $P$ e $NP$.
    Come \`e noto questo \`e uno dei cosiddetti problemi del Millennio:
    non ci si aspetta che ammetta una soluzione semplice e verificarne
    la correttezza pu\`o essere difficile e dispendioso.
    Nessuno dei lavori contenuti nella lista di Woeginger
    \`e stato pubblicato
    su riviste peer reviewed e/o di elevata reputazione scientifica. Tuttavia, gli autori non sono degli sprovveduti, ma ricercatori di notevole esperienza, e
    frequentemente docenti universitari della materia in oggetto o di settori
    affini.

    Sarebbe bello poter disporre di uno strumento automatico a cui dare in pasto
    questi articoli, e che rispondesse asserendo la correttezza o meno
    delle argomentazioni addotte. Questo \`e profondamente diverso
    dal cercare di {\em dimostrare automaticamente} l'uguaglianza $P=NP$.
    Se abbiamo un sistema
    logico con un minimo di espressivit\`a la {\em ricerca} di una dimostrazione
    non \`e un problema decidibile, ma il problema di determinare se un termine
    di prova $p$ \`e sintatticamente corretto, e in tal caso se rispetta un
    determinato enunciato $P$ (useremo la notazione $p:P$ per esprimere che
    $p$ \`e una dimostrazione di $P$) \`e non solo decidibile, ma anche relativamente
    semplice, almeno da un punto di vista concettuale\footnote{La complessit\`a pratica del problema dipende molto dal sistema logico di riferimento e dalla nozione di prova}:
    si tratta di seguire passo passo la dimostrazione e verificare {\em localmente}
    che non siano stati fatti errori nell'applicazione delle regole logiche.
    Il problema della dimostrazione automatica \`e in effetti un problema di
    {\em ricerca} nello spazio delle prove, che {\em presuppone} la possibilit\`a
    di {\em verificare} che la prova sia corretta.
    
    \vspace{.3cm}
    \begin{tabular}{|c|c|}
      \hline
      {\bf Verifica Automatica}                  & {\bf Dimostrazione Automatica}\\
      $p$ \`e una dimostrazione di $P$ ?   & esiste $p$ che dimostra $P$ ?\\\hline
    \end{tabular}
    \vspace{.3cm}
    
    La dicotomia ricerca vs. verifica \`e uno dei leit-motif
    dell'informatica: verificare pu\`o essere semplice, mentre cercare \`e,
    di solito, molto pi\`u complicato. 
    Per esempio, i problemi in $NP$ sono quei problemi decisionali che ammettono un
    algoritmo efficiente (polinomiale) di {\em verifica} della risposta
    per mezzo di un certificato di dimensione polinomiale nella dimensione dell'input.
    La ricerca di una soluzione mediante tecniche di generate-and-test che
    generano in modo esaustivo tutti i potenziali certificati e ne {\em verificano}
    la correttezza, fornisce tipicamente algoritmi deterministici
    di complessit\`a esponenziale. Tuttavia, molti problemi in $NP$ hanno anche
    degli algoritmi ad hoc di {\em ricerca} efficienti: questi sono i problemi in $P$. 

    Se dunque si ammette che la verifica sia un problema relativamente semplice
    (almeno rapportato alla dimostrazione automatica)
    che cosa rende interessante questa area di ricerca e complicati i relativi
    strumenti di elaborazione?

    La complessit\`a deriva non dal tool di verifica, che \`e una piccola componente
    di solito chiamata ``kernel'' (si veda ad esempio \cite{ck-sadhana}), ma
    dall'{\em ambiente di sviluppo} fornito
    all'utente per lo sviluppo assistito e interattivo delle dimostrazioni \cite{matita-crafting}.
    \`E questo ambiente che fa si che, mediante l'uso di proof assistants, si possano
    certificare prove di notevole difficolt\`a, come il teorema di Feit-Thompson sui
    gruppi finiti \cite{GonthierAll13}, o garantire la correttezza di componenti
    software particolarmente
    complessi e delicati, come compilatori \cite{Leroy09},
    parti di sistemi operativi \cite{Klein10}, sistemi di
    controllo per applicazioni critiche \cite{krakatoa} o strumentazioni hardware \cite{Itanium}.
    
    La prospettiva rispetto alla deduzione automatica \`e a questo punto completamente
    rovesciata. L'ambiente di sviluppo \`e uno strumento che permette di
    costruire dimostrazioni complesse a partire da dimostrazioni pi\`u semplici sotto
    la guida dell'utente. Parte delle dimostrazioni pi\`u semplici, o parte di quelle
    manipolazioni logico-algebriche che semplificano la combinazione dei risultati (ad esempio,
    adattamento di strutture a meno di semplici isomorfismi o istanziazioni),
    possono essere delegate
    a strumenti automatici di ricerca. Il dimostratore interattivo permette dunque di
    combinare svariate tecniche di prova (di solito chiamate tattiche), tra cui {\em anche} tecniche
    di dimostrazione automatica. Mentre la verifica automatica \`e un ovvio
    presupposto per la deduzione automatica, quest'ultima \`e solo una componente
    nel variegato campo della dimostrazione interattiva.
    \begin{center}
      \begin{figure}[htb]
      \includegraphics[width=.8\textwidth]{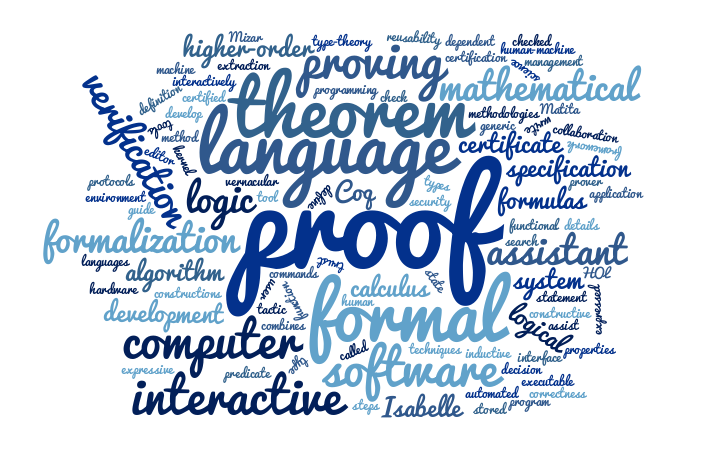}
      \caption{L'universo della dimostrazione assistita. La nuvola di parole
        \`e stata calcolata in base al contenuto di alcuni siti web
        specializzati su questo argomento.}
      \end{figure}
    \end{center}
    
    \section{Dimostrazione formale}

    Torniamo al caso dell'autore della dimostrazione di $P=NP$, probabilmente
    frustrato dalla
    scarsa considerazione ottenuta nell'ambiente scientifico dal proprio risultato.
    Quello che potrebbe
    fare, con uno sforzo sicuramente comparabile alla rilevanza del problema,
    sarebbe di formalizzare il risultato in uno degli svariati tool di dimostrazione
    interattiva esistenti e autenticare formalmente la correttezza della propria
    argomentazione.

    Questo ci porta al problema cruciale: cosa intendiamo con dimostrazione formale
    e quanto \`e difficile passare da una dimostrazione ``cartacea'' ad una
    completamente formale e verificabile tramite un computer?

    Una dimostrazione formale \`e
    semplicemente una dimostrazione che possa essere letta ed elaborata da
    una macchina. Il linguaggio matematico \`e spesso considerato rigoroso,
    ma non ha quella turgidit\`a sintattica propria dei linguaggi formali, ed
    in particolare dei linguaggi di programmazione, che semplifica enormemente
    la comunicazione tra uomo e macchina. Questo \`e il primo passo importante:
    si deve accettare l'idea di riscrivere la dimostrazione in un linguaggio
    formale, e che l'apprendimento di questo linguaggio potrebbe richiedere
    un certo sforzo (comparabile, indicativamente, allo sforzo di apprendere un
    nuovo linguaggio di programmazione).

    Questo spiega anche la presunta cripticit\`a delle dimostrazioni
    formali: il linguaggio delle prove di queste applicazioni non \`e inteso
    per la comunicazione inter-umana, ma per la comunicazione tra l'uomo
    e la macchina. Un utente esperto \`e in grado di interpretare questo
    linguaggio, possibilmente in un dialogo interattivo con la macchina,
    ma il significato rester\`a essenzialmente opaco ai non addetti ai
    lavori. Allo stesso modo, in informatica, l'algoritmo \`e celato
    dietro al programma, e chi non conosce il particolare linguaggio di
    programmazione in uso avr\`a inevitabilmente difficolt\`a a interpretare il
    significato di un certo codice e ricostruirne l'algoritmo.

    La dimostrazione matematica sta alla sua controparte formale esattamente
    come l'algoritmo sta ad un programma che lo implementa.
    
    Il problema \`e a volte presentato in relazione al duplice ruolo delle
    dimostrazioni: quello di {\em certificato} e quello di {\em messaggio}.
    \`E stata recentemente avanzata l'ipotesi \cite{science} che l'avvento dei
    dimostratori automatici conduca a un sostanziale divorzio di questi due
    ruoli (non \`e chiaro se l'autore interpretasse questo fenomeno in senso positivo
    o negativo). In realt\`a, se da un lato \`e indubbiamente interessante
    separare
    queste due funzioni della nozione di prova, \`e altrettanto importante
    mantenere un {\em allineamento} tra di esse, problematica sicuramente
    centrale allo sviluppo del settore della dimostrazione interattiva.

    Per meglio approfondire questa problematica, dobbiamo fare riferimento
    ai due principali stili di scrittura delle dimostrazioni utilizzati
    dai dimostratori interattivi, quello procedurale e quello dichiarativo,
    che andiamo ad analizzare nella prossima sezione.
    
    \subsection{Stile dichiarativo e stile procedurale}

    Una dimostrazione \`e un procedimento (giustificazione)
    che permette di ricavare una
    conclusione $C$ a partire da un insieme di premesse $A_1,\dots A_n$.
    La dimostrazione pu\`o tipicamente essere decomposta in una sequenza
    di passi pi\`u semplici, fino ad arrivare a un insieme di passi
    elementari non ulteriormente scomponibili, che costituiscono le
    regole logiche fondamentali del ragionamento deduttivo, caratteristiche
    del particolare sistema logico adottato. Le regole logiche sono
    casi particolari di dimostrazioni la cui validit\`a viene assunta
    assiomaticamente e dunque non richiedono ulteriore giustificazione se
    non la semplice menzione della regola invocata.
    Ad esempio, una regola di {\em introduzione della congiunzione},
    chiamiamola $\wedge_I$, permette di concludere $A \wedge B$ a
    partire dalle premesse $A$ e $B$:
    \[
    \wedge_I: \frac{A \hspace{1cm} B}{A \wedge B}
    \]
    Dovrebbe essere chiaro che c'\`e una certa ridondanza nella
    nozione suddetta di regola logica: conoscendo premesse e conclusione
    esiste abitualmente una sola regola logica che giustifica il
    passo deduttivo, facilmente individuabile dalla macchina;
    inoltre, e questo \`e molto pi\`u interessante,
    conoscendo il nome della regola applicata si conosce la relazione
    strutturale tra premesse e conclusione, e quindi si possono inferire
    automaticamente le une dalle altre.

    Questo ci porta alla definizione delle due classi principali di
    linguaggi di specifica di dimostrazioni: dichiarativi o procedurali.
    Nell'accezione di Hilbert, una dimostrazione di $P_n$, \`e una
    sequenza di formule $P_1,\dots,P_n$ dove ogni $P_i$ \`e una
    ipotesi oppure si ottiene da formule che precedono $P_i$ nella
    sequenza per mezzo di una specificata regola logica. Tuttavia, come
    spiegato precedentemente, per definire la
    prova \`e sufficiente fornire la sequenza delle regole applicate
    (descrizione procedurale) lasciando implicita la sequenza delle
    formule, oppure fornire la sequenza $P_1,\dots,P_n$ dei passi
    intermedi delegando al sistema il semplice compito di inferire
    per conto proprio le regole logiche applicate nei singoli passi
    (descrizione dichiarativa).

    La differenza tra linguaggi procedurali e 
    dichiarativi e i vantaggi degli uni rispetto agli altri possono
    essere capiti meglio con una semplice analogia
    con il gioco degli scacchi (ripresa da \cite{message}).

\begin{center}
      \begin{figure}[htb]
      \includegraphics[width=1.\textwidth]{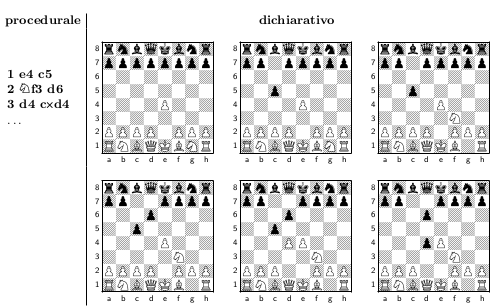}
      \caption{Mosse (stile procedurale) e Posizioni (stile dichiarativo)}
      \label{Fig:chess1}
      \end{figure}
    \end{center}

    Una partita di scacchi pu\`o essere descritta essenzialmente
    in due modi: mediante una sequenza di {\em mosse} o mediante
    una sequenza di {\em posizioni} che descrivono lo stato dei
    pezzi sulla scacchiera (si veda la Figura~\ref{Fig:chess1}).

    Nel primo caso, le posizioni sono implicite: esse possono essere
    ricostruite in modo univoco a partire dalla sequenza delle
    mosse. Nel secondo caso, le mosse sono implicite, e possono
    essere dedotte dalla differenza tra due posizioni successive.
    Mosse e posizioni sono rispettivamente esempi di un linguaggio
    procedurale e di uno dichiarativo.

I meriti relativi dei due approcci dovrebbero essere evidenti.
La descrizione procedurale \`e {\em estremamente} compatta ma piuttosto
illeggibile: ogni mossa fa implicito riferimento alla posizione
raggiunta mediante le mosse precedenti. Per seguire una descrizione
procedurale bisognerebbe essere in grado di immaginare le configurazioni
intermedie risultanti dall'esecuzione di ogni singolo passo.
Nel caso degli scacchi i giocatori esperti riescono abitualmente
a farlo, visto che le mosse sono operazioni abbastanza elementari e
le partite non sono mai troppo lunghe. Tuttavia, nel caso delle
dimostrazioni, questo \`e molto pi\`u complicato. Nella maggior
parte dei casi, le dimostrazioni in stile procedurale sono praticamente
inintelleggibili per l'uomo: il punto cruciale \`e che non sono
pensate per essere {\em lette}, ma per essere {\em replicate}
interattivamente al computer (cosi come si pu\`o sviluppare la partita
di scacchi su una scacchiera eseguendo passo passo le mosse descritte
a partire dalla configurazione iniziale).

D'altra parte, le descrizioni procedurali forniscono, ad ogni istante,
una descrizione esplicita e completa dello stato corrente. L'evoluzione
futura non dipende dal passato, e ci si pu\`o concentrare interamente
sul presente. Le descrizioni dichiarative sono dunque tipicamente pi\`u
leggibili di quelle procedurali, ma anche molto pi\`u prolisse, al
punto da diventare spesso tediose per l'utente.

\subsection{Un esempio concreto}
Per confrontare i due metodi facciamo un semplice esempio di
applicazione dell'analisi per casi sui numeri naturali (una forma degenere
dello schema di induzione):

\[(*)\;\; \frac{P(0) \hspace{.5cm}\forall x.P(S(x))}{P(t)}\]

Supponiamo che
l'obiettivo (il nostro goal corrente $G$) sia dimostrare che ogni numero naturale $n$ diverso da
zero \`e il successore di un qualche naturale $m$

\[(G)\;\;n \neq 0 \to \exists m, S(m) = n\]

Si vuole procedere per casi su $n$, intenzione
che, in un linguaggio procedurale, si traduce tipicamente nel semplice
comando
\[\mbox{cases } n\]
Innanzitutto, il sistema capisce dal tipo di $n$ il tipo di analisi
da effettuare, e che in questo caso il principio da utilizzare \`e $(*)$.
Tuttavia, $(*)$ \`e uno {\em schema}, per cui il problema successivo
\`e capire quale \`e la propriet\`a $P$ a cui si fa riferimento.
Questa viene inferita in modo automatico a partire da $G$ astraendo
rispetto alle occorrenze sintattiche del termine $n$ (che potrebbe
essere un termine complesso, non necessariamente una variabile).
Il predicato che si ottiene \`e:
\[P = \lambda n.n \neq 0 \to \exists m, S(m) = n\]
Utilizzando questo predicato, il sistema ricostruisce per semplice
istanziazione le due ipotesi corrispondenti alle premesse di (*),
riducendo quindi $G$ ai due sottogoal seguenti:
\[
\begin{array}{l}
(G_1)\;\; (0 \neq 0 \to \exists m, S(m) = 0\\
  (G_2)\;\; \forall x, S(x) \neq 0 \to \exists m, S(m) = S(x)
\end{array}
\]
che \`e quanto ci aspettavamo.

Questo passo di dimostrazione in linguaggio procedurale \`e dunque
interamente racchiuso nel comando ``cases $n$'', dove $n$ fa {\em implicitamente}
riferimento al goal corrente $G$. La stessa dimostrazione
in linguaggio dichiarativo \`e descritta dalla sequenza dei goals $G_1,G_2,G$,
possibilmente decorati con del testo aggiuntivo, sia per ragioni di
leggibilit\`a che per semplificare il lavoro di verifica. Ad esempio:
\[
\begin{array}{l}
(G_1)\;\; (0 \neq 0 \to \exists m, S(m) = 0\\
 (G_2)\;\; \forall x, S(x) \neq 0 \to \exists m, S(m) = S(x)\\
  \mbox{By case analysis from $G_1$ and $G_2$ we obtain}\\
  (G)\;\;n \neq 0 \to \exists m, S(m) = n
\end{array}
\]

\subsection{Breve storia}
I sistemi di prova procedurali discendono da
uno strumento dei primi anni
settanta ideato dal premio Turing Robin Milner e denominato LCF \cite{lcf},
un acronimo per Logic of Computable Functionals (si veda anche
la sottosezione \ref{sec:HOL}, pi\`u avanti). Tali sistemi hanno
una nozione di ``stato corrente della prova'' che consiste in un
insieme di cosiddette ``proof obligations'' corrispondenti ai
sottogoal pendenti che ancora devono essere dimostrati. Lo stato
\`e trasformato mediante le cosiddette {\em tattiche} che prendono
un goal e lo riducono a un insieme (possibilmente vuoto)
di sottogoal pi\`u semplici. La dimostrazione comincia con l'asserto
da dimostrare come unica obbligazione e termina quando non resta
nessun sottogoal aperto. A questo punto la dimostrazione pu\`o essere
salvata in memoria e diventa utilizzabile per usi futuri. Per via
del processo di prova, che {\em parte} dall'asserto finale e procede
riducendo via via goal complessi a goal pi\`u semplici, le dimostrazioni
procedurali vengono tipicamente costruite {\em all'indietro} (backward),
dalla conclusione verso le ipotesi.

Il primo linguaggio ad adottare uno stile dichiarativo \`e stato
il sistema di prova Mizar \cite{mizar85,mizarJFR}, sviluppato fin dai
primi anni settanta
dal logico polacco Trybulec, recentemente scomparso. Il linguaggio
di Mizar \`e stato espressamente studiato per essere comprensibile
all'uomo, pur mantenendo un livello di formalit\`a sufficiente per
essere decifrabile dalla macchina.

Mizar era poco noto in occidente. In seguito ad una sua visita
al gruppo di Trybulec in Bialystok, J.Harrison, sviluppatore di
HOL-light, prov\`o a ideare uno stile di prova procedurale per il
proprio proof assistant, detto appunto Mizar-mode \cite{harrison-mizar},
che contribu\`i in modo notevole a popolarizzare questo stile di
scrittura delle dimostrazioni. Molti altri strumenti nati in forma
procedurale hanno in seguito sviluppato intefacce in grado di gestire
prove scritte in forma dichiarativa: si veda ad
esempio \cite{declarative-coq} per COQ, \cite{csc-decl} per Matita,
o il linguaggio Isar \cite{Isar} per Isabelle. Quest'ultimo \`e
probabilmente, al momento attuale, l'esempio pi\`u compiuto di linguaggio
dichiarativo, ed ha praticamente soppiantato lo stile procedurale nell'ambito
degli utenti di Isabelle. Si vedano anche i seguenti lavori per un interessante
raffronto fra i vari linguaggi: \cite{WenzelW02,harrison-proof_style}.

Come dovrebbe essere chiaro dalla descrizione che abbiamo fatto e
dai risultati precedenti, la distanza tra linguaggi procedurali e
dichiarativi non \`e cosi ampia come potrebbe apparire a prima vista:
in realt\`a sono complementari
uno all'altro e si integrano bene insieme \cite{Mizar-HOL,KaliszykW08,Freek-synthesis}.
Da uno script procedurale
\`e possibile di solito ricostruire in modo automatico
una versione dicharativa apprezzabile, cos\`i come, in linea di principio,
sarebbe possibile riassumere una descrizione procedurale in forma
dicharativa (poco esplorato, e forse poco utile).

Tuttavia, pi\`u che una integrazione tra i due stili, l'obiettivo
probabilmente pi\`u importante per l'approccio procedurale \`e quello
di arrivare ad una descrizione {\em maggiormente strutturata}
delle dimostrazioni, mirata non tanto ad incrementarne la leggibilit\`a,
quanto la loro manutenzione e il loro aggiornamento. 
Questa tendenza \`e chiaramente testificata dai tool pi\`u
recenti come Ssreflect \cite{ssreflect} e Matita \cite{matita-cade,matita-tutorial}.

\subsection{Obiettivi comuni, strategie contrastanti}
L'obiettivo comune, sia per l'approccio procedurale che per quello dichiarativo
\`e un progressivo {\em aumento della granularit\`a della dimostrazione}, il che
corrisponde essenzialmente ad un potenziamento delle regole logiche.

Nel caso dell'approccio procedurale, la regola logica viene inferita in modo
automatico dal sistema, quindi l'obiettivo \`e un {\em aumento delle sue capacit\`a di
dimostrazione automatica}, o in alternativa la sua interoperabilit\`a con sistemi
esterni cui delegare la sintesi di questi passi di dimostrazione.
La tematica \`e cosi descritta in \cite{Hammering16}:
\begin{quote}
  The main ingredients underlying this approach are efficient automatic theorem provers that can cope with hundreds of axioms, suitable translations of richer logics to their formalisms, heuristic and learning methods that select relevant facts from large libraries, and methods that reconstruct the automatically found proofs inside the proof assistants.
\end{quote}
Lo scopo finale \`e quello di descrivere la prova come un elenco (possibilmente strutturato)
di un certo numero di passi intermedi particolarmente significativi, eventualmente
inframezzato da annotazioni che aiutino l'automazione a sintetizzare la prova.
Queste annotazioni protrebbero essere semplicemente un elenco dei risultati utili
a completare la prova: ``otteniamo $P$ mediante $H_1,\dots H_n$''. Queste permette
al dimostratore automatico di confinare lo spazio di ricerca all'interno delle
librerie dei risultati noti, che potrebbe altrimenti risultare troppo ampio.
Questo settore \`e al  momento attuale uno dei pi\`u vivaci e promettenti,
con particolare riferimento all'uso di tecniche di apprendimento automatico per
migliore le prestazioni dei dimostratori \cite{Urban15,KU15}.

Nel caso dell'approccio dichiarativo, i passi di prova sono le tattiche, ovvero
i comandi che l'utente utilizza per procedere nella dimostrazione e ridurre un goal
a goal pi\`u semplici. Aumentare la granularit\`a significa concepire tattiche
via via pi\`u espressive, potenti e generali. Dei particolari costrutti, detti
{\em tatticali} permettono di combinare assieme tattiche per generare comandi ancora
pi\`u espressivi. Determinati sistemi permettono inoltre agli utenti di definire
le proprie tattiche, fornendo della applicative user interfaces per la manipolazione
dei termini di prova. In molti casi, questi linguaggi sono utilizzati per costruire
dei dimostratori automatici {\em specifici per un determinato dominio}, in contrasto
con le tecniche generali dell'approccio procedurale.

Un'altra importante tematica di sviluppo dei linguaggi procedurali \`e il progressivo
uso della {\em riflessione} \cite{Ring,BarendregtB02} per ridurre passi logici a semplici problemi computazionali.
Per fare un esempio, supponiamo che l'uguaglianza tra elementi di un certo tipo $A$
sia decidibile, cio\`e che si disponga di un algoritmo di decisione $eq_A:A\to A \to bool$
e supponiamo di avere dimostrato la sua correttezza, ovvero che
\[\forall x,y:A, eq_A(x,y) = true \Leftrightarrow x = y\]
Allora possiamo {\em riflettere} la dimostrazione di uguaglianza tra due elementi $a$ e $b$
di tipo $A$ sul semplice calcolo di $eq_A(a,b)$.
Questo \`e un tipico esempio di small scale reflection, promosso da \cite{ssreflect}. 
Qualunque predicato decidibile da luogo a una forma interessante di riflessione.

Le complesse tattiche di prova correntemente utilizzate nei dimostratori
procedurali sembrano difficilmente rimpiazzabili da procedure automatiche
generaliste. Nonostante l'innegabile appealing dello stile dichiarativo
non \`e al momento immaginabile che questo possa completamente soppiantare
lo stile procedurale, soprattutto nell'ambito delle applicazione
all'informatica (verifica di propriet\`a di componenti software o hardware).

\section{Il quadro fondazionale}
Il sistema logico fondazionale utilizzato
dal sistema interattivo di dimostrazione ha un ruolo pi\`u marginale
di quanto si possa immaginare. 
La ragione di questo fatto \`e che tra l'utente e il livello fondazionale
possono essere aggiunti svariati livelli intermedi che permettono di
lavorare ad un livello di astrazione maggiore rispetto al formalismo
inevitabilmente grezzo, pedante e prolisso fornito dal sistema logico
di base. 

Per fare una analogia con i linguaggi di programmazione, il paradigma
fondazionale fornisce un linguaggio di basso livello, atto a essere eseguito
su strutture hardware, ma di difficile utilizzo da parte dell'uomo.
In informatica, i sistemi operativi permettono di astrarre dall'architettura
fisica del sistema, virtualizzando i suoi componenti, e i compilatori
permettono di astrarre dal linguaggio
macchina di basso livello, consentendo all'utente di scrivere programmi in
linguaggi molto pi\`u evoluti.
Allo stesso modo, ogni proof assistant fornisce un linguaggio di
prova ``di alto livello'' e un complesso sistema
di astrazioni che tendono a sopperire alle inevitabili lacune e
restrizioni del quadro fondazionale, e al tempo stesso diminuire
la dipendenza da esso.

\vspace{.4cm}
\begin{center}
    \begin{tabular}{|c|}
      \hline
      hardware = sistema logico fondazionale\\
      \hline
    \end{tabular}
    \end{center}
    \vspace{.4cm}
    

L'osservazione precedente fa cadere l'obiezione principale abitualmente
addotta per sostenere l'impossibilit\`a {\em pratica} di un approccio
strettamente formale alla matematica (si veda ad esempio \cite{DeMillo79}),
derivante dall'eccessiva lunghezza delle dimostrazioni in oggetto.
Bourbaki stesso, pur essendo considerato un fautore
di un approccio maggiormente formale alla matematica \cite{lee}, riteneva
che il progetto di una completa formalizzazione di questa disciplina fosse
{\em assolutamente irrealizzabile}, e non per ragioni di incompletezza ma
di pura dimensione delle dimostrazioni \cite{bourbaki}:

\begin{quotation}
\em
the tiniest proof at the beginning of the Theory of Sets would already
require several hundreds of signs for its complete formalization.
\end{quotation}
Come osservato in \cite{social}, l'argomentazione ricorda la generale
mancanza di fiducia nella
possibilit\`a di scrivere programmi per svolgere compiti complessi
tipica della della prima met\`a degli anni cinquanta, nel periodo
antecedente alla costruzione dei primi compilatori e all'avvento
dei linguaggi di programmazione di alto livello.
Per citare Harrison \cite{NAMS-Harrison}:
\begin{quotation}
\em
the arrival of the computer changes the situation dramatically.
[...] checking conformance to formal rules is one of the things computers
are very good at. [...] the Bourbaki claim that the transition to a 
completely formal text is routine seems almost an open invitation
to give the task to computers.
\end{quotation}

Allo stato attuale dell'arte, la presunta lunghezza delle dimostrazioni
formali non rappresenta pi\`u un ostacolo, come ampiamente
dimostrato dalla completa formalizzazione e verifica automatica
della correttezza di risultati matematici complessi come 
la distribuzione asintotica dei numeri primi \cite{Avigad07}, 
il teorema dei quattro colori \cite{NAMS-Gonthier}, il
teorema della curva di Jordan \cite{Hales-jordan} o il gi\`a citato
teorema di Feit-Thompson sui gruppi finiti \cite{GonthierAll13}.

\subsection{Teoria degli insiemi vs. Teoria dei tipi}
Bench\`e per l'utente finale il quadro fondazionale non
sia (o non dovrebbe essere) cos\`i rilevante, \`e comunque interessante,
anche per ragioni
storiche, fare un piccolo escursus sui sistemi logici tipicamente adottati
dai dimostratori interattivi. Questo ci permetter\`a anche di
discutere e capire meglio l'architettura complessiva dei diversi sistemi.

\`E opinione diffusa che il ``vero'' fondamento della matematica sia la
teoria degli insiemi, tuttavia non esiste una ragione profonda a supporto
di questa tesi, ed \`e significativo osservare che la quasi totalit\`a
dei sistemi di prova interattivi, con l'unica eccezione di Mizar,
{\em non si basano su questa teoria fondazionale}.

Le ragioni di questo fatto non sono facilmenti individuabili e sicuramente
ci sono anche delle importanti motivazioni storiche, dovute al particolare
momento in cui questi sistemi sono stati sviluppati,
alle tematiche che si stavano approfondendo, e ai gruppi di ricerca a cui
appartenevano gli sviluppatori, o di cui subivano l'influenza.

%
%

Volendo cercare una motivazione generale che rende la teoria degli insiemi
poco attraente dal punto di vista della verifica formale, il punto
nodale riguarda probabilmente il concetto di funzione. In teoria
degli insiemi una funzione non \`e che un caso particolare di
relazione, totale e univoca. Questo \`e in apparente contrasto
con la centralit\`a della nozione di funzione, sia in matematica
che, forse ancor pi\`u, in informatica (dove, ai fini della verifica,
la funzione \`e l'oggetto principale del discorso).
\`E naturale voler disporre
di un framework logico-linguistico di base che permetta la definizione
e la manipolazione pi\`u semplice possibile di queste entit\`a. 

Una caratteristica comune a tutti i moderni sistemi di prova
interattivi \`e l'utilizzo di logiche di ordine superiore.
I vari sistemi
si differenziano tuttavia per le forme variegate che possono assumere
questi sistemi logici, dalla teoria
dei tipi semplici di Church, ai vari sistemi intuizionisti, predicativi
o impredicativi, con sottotipi, tipi dipendenti, e via dicendo.
Questa pletora di sistemi, tanto simili uno all'altro e tuttavia
praticamente incompatibili tra loro \`e una delle principali
ricchezze/disgrazie di questa area di ricerca.

\subsection{LCF e la famiglia HOL}
\label{sec:HOL}
Un primo gruppo di dimostratori \`e quello che appartiene alla famiglia
HOL (si veda \cite{Gordon00,Gordon08} per delle interessanti rassegne
storiche). I sistemi di questa famiglia si isipirano alla Logic of Computable
Functionals (LCF) di Milner \cite{lcf}, che racchiude il sistema logico
all'interno
di un metalinguaggio programmabile (nel caso di Milner, il linguaggio ML).
I termini e le formule sono espressioni del linguaggio, che possono essere
manipolate liberamente. I teoremi sono valori di un particolare {\em tipo di
  dato astratto}, i cui metodi sono le regole di base del sistema logico.
Fintanto che solo queste funzioni sono utilizzate per costruire nuovi
teoremi (cosa che pu\`o essere verificata dal sistema dei tipi del
metalinguaggio di programmazione) si ha la garanzia di correttezza delle
deduzioni fatte. Questo per quanto riguarda la costruzione in avanti
delle dimostrazioni. Per supportare il ragionamento all'indietro, si
implementano goals e tattiche sul tipo di dato astratto. Ogni tattica
verifica che il goal in input ha la forma sintattica che ci si attende,
genera i nuovi sottogoals e sintetizza la funzione di validazione
che giustifica la conclusione a partire dalle premesse. Funzioni di
validazione e tattiche sono dunque essenzialmente inverse le une alle
altre. La dimostrazione (validazione) non \`e di per s\`e un oggetto
sintattico del linguaggio, ma un programma: pu\`o essere eseguito,
ma non ispezionato.
Le tattiche possono essere combinate assieme mediante dei costrutti
pi\`u complessi, detti tatticali, che consentono di implementare passi logici
di notevole complessit\`a. L'utente \`e incoraggiato a scrivere le proprie
tattiche e i propri tatticali, spesso caratteristici del particolare ambito
in cui si sta lavorando.

Ogni sistema logico pu\`o essere implementato in questa forma, ma la
famiglia HOL si caratterizza appunto dal fatto di adottare come paradigma
fondazionale varianti della logica di ordine superiore basate sulla teoria
dei tipi semplici di Church.

\subsection{Isabelle}
Isabelle \`e un framework logico originariamente sviluppato da Paulson
a Cambridge verso la fine degli anni 80. La concezione di Isabelle
deriva dallo sforzo di aggiungere al modello architetturale di LCF una
visione maggiormente integrata della dimostrazione. In particolare, i
sottogoal generati da una tattica possono condividere informazioni
(ad esempio termini non ancora instanziati) richiedendo una nozione di
metavariabile ``esistenziale'' e la gestione delle operazioni rilevanti
associate a questa, a partire dalla nozione di unificazione.
La conseguenza \`e quella di spostare l'attenzione dalla nozione di
teorema (centrale all'approccio LCF) verso quella di sequente, in
particolare in relazione alle regole di inferenza. 

Mentre nell'approccio LCF ogni regola di inferenza \`e implicitamente
rappresentata da una funzione
per la sua applicazione in avanti (validazione) e da una per la sua
applicazione all'indietro (tattica), in Isabelle, ogni regola di inferenza
\`e descritta da una clausola di Horn. Le regole logiche si compongono
(istanziandosi) mediante il meccanismo di risoluzione, senza bisogno
di differenziare tra il loro uso in avanti o all'indietro.

Un sistema di ordine superiore minimale, chiamato $M$,
\`e utilizzato come framework
sintattico in cui esplicitare le regole. Isabelle \`e un esempio di
Logical Framework: per poter utilizzare lo strumento concretamente
bisogna prima istanzianziarlo su una qualche logica pi\`u espressiva
e quindi lavorare all'interno di questa (detta logica oggetto, per
distinguerla dalla meta-logica di Isabelle). La versione attualmente
pi\`u utilizzata di Isabelle \`e Isabelle-HOL, che utilizza HOL come
logica oggetto, estesa con un nucleo assiomatico per permettere
ragionamenti di tipo set-theoretic; esistono per\`o delle librerie
per constructive type theory (CTT) e Zermelo Fraenkel Set Theory.

Per concludere, la differenza tra Isabelle e HOL pu\`o essere
essenzialmente riassunta nella seguente frase di Paulson \cite{Paulson89}:

\begin{quote}I have criticized LCF for representing inference rules
  as functions, preferring visible structures. But colleagues have pointed
  out that such structures are essentially the theorems of a (meta) logic,
  and the functions that manipulate them are (meta) rules. Thus Isabelle
  adopts LCF's representation at the meta-level.
\end{quote}

\subsection{Teoria dei Tipi}
L'altro grande filone di dimostratori interattivi \`e quello che si
ispira alla teoria dei tipi, ed in particolare alle cosiddette
teorie costruttive. Appartengono a questa categoria Coq \cite{CoqArt},
Agda \cite{agda}, Matita \cite{matita-tutorial},
e bench\`e in forma leggermente differente, Nuprl \cite{nuprl-book}.

Queste applicazioni sono il risultato di un fecondo lavoro di ricerca
che a partire dagli anni settanta ha coinvolto, in strettissima
sinergia tra di loro, teoria delle categorie, costruttivismo e
lambda calcolo (si vedano ad esempio
\cite{martinloef84,lambek88,GirardJY:prot} per una introduzione all'argomento).
Questa relazione \`e tutt'altro che esaurita, come
dimostrano le recenti scoperte nel campo della Homotopy Type Theory
(HoTT) \cite{hottbook} promossa da Voevodsky, vincitore di una medaglia
Field per i suoi lavori sulle coomologie algebriche.

Dal punto di vista della verifica formale, i punti salienti dell'approccio
type-theoretic sono essenzialmente due: la nozione di tipo induttivo
e la cosiddetta analogia di Curry-Howard.

\subsubsection{Tipi induttivi}
Col termine {\em tipi induttivi} si intende un meccanismo uniforme
ed estremamente potente che permette
di definire tipi di dato ricorsivi {\em unitamente} ai rispettivi
principi di eliminazione, ovvero, in termini di programmazione,
unitamente ai meccanismi di ricorsione che permettono la visita e
la elaborazione strutturata dei dati in oggetto.

Questo \`e un punto filosofico caratterizzante della teoria dei tipi
costruttiva.
Il tipo non \`e semplicemente una collezione di oggetti con una certa
etichetta, ma \`e definito dai suoi {\em costruttori}, dai suoi
{\em eliminatori} (distruttori) e dalle regole di interazione tra costruttori
e distruttori (regole di riduzione). Quindi un tipo non \`e una collezione
di oggetti statici, ma definisce anche la {\em dinamica} dei propri elementi,
fornendo una procedura di {\em normalizzazione} degli stessi. Non si
ha dunque solo una logica per ragionare sui dati, ma anche un calcolo
per esprimere computazioni su di essi.

\subsubsection{Analogia di Curry-Howard}
L'analogia di Curry-Howard \cite{curry-howard} \`e una corrispondenza formale tra logica e
programmazione che mette in relazione le formule logiche e le relative
dimostrazioni da un lato, con i tipi e i programmi dall'altro.
La relazione \`e in stretta relazione con la intrepretazione operazionale
di Brouwer-Heyting-Kolmogorov, per cui una dimostrazione di $A_1\dots A_n \to B$
\`e un procedimento costruttivo di sintesi di una prova di $B$ a partire
dalle prove di $A_1,\dots, A_n$. Con isomorfismo di Curry-Howard si intende
una versione formale di questa corrispondenza tra vari sistemi di
logica intuizionista e opportune versioni del lambda calcolo tipato
(\cite{berardi-pts,BarendregtH:lawcwt,SU06}).
In accordo a questa analogia, verificare che una dimostrazione \`e corretta
equivale a verificare che un programma \`e ben tipato
(il cosiddetto type-checking), che \`e la
principale operazione di analisi statica dei programmi svolta dai compilatori.
Al tempo stesso, l'analogia permette di avere notazioni e strutture dati
uniformi per prove e programmi, semplificando sensibilmente la struttura
architetturale dei sistemi di prova.
\begin{table}
\caption{Corrispondenza di Curry-Howard}
\begin{tabular}{ll}
 \hline \hline
{\em Matematica}\hspace{2.0cm} & {\em Programmazione}\\
 \hline
teorema & tipo\\
prova & programma \\
verifica di correttezza & type checking \\
eliminazione dei tagli & computazione \\
 \hline \hline
\end{tabular}
\label{curry-howard}
\end{table}
La corrispondenza di Curry-Howard apre inoltre una prospettiva
completamente nuova sulla verifica dei programmi: invece
di cercare di stabilire l'adeguatezza di un programma rispetto alla sua
specifica, si pu\`o cercare semplicemente di dimostrare la coerenza
logica della specifica. Se la dimostrazione \`e costruttiva, \`e
possibile estrarre in modo automatico il contenuto algoritmico della
dimostrazione, ottenendo un programma che \`e garantito essere corretto
per costruzione. Questo procedimento, noto come {\em estrazione di programmi},
\`e stato sfruttato per la prima volta nel sistema 
Nuprl \cite{nuprl-book}, ed \`e in seguito stato implementato
in molti altri sistemi \cite{Moh89a,letouzey02}. L'analogia di Curry-Howard e la tecnica
di estrazione possono essere anche estesi, entro certi limiti, ai
sistemi logici classici \cite{parigot92,berardi96}.

\subsection{Altri dimostratori interattivi}
Esistono svariati altri importanti dimostratori interattivi
che non rientrano nelle due grandi famiglie precedenti. Questi si
differenziano per i meccanismi logici leggermente diversi che sono
supportati, il linguaggio di programmazione in cui sono scritti,
e le finalit\`a per cui sono stati progettati. Tra questi ricordiamo
in particolare PVS \cite{pvs92}, Twelf \cite{twelf} e il recente
sistema Dedukti \cite{dedukti} basato sul lambda-pi-calcolo modulo
di Dowek \cite{lambda-pi-modulo,modulo}. Altri sistemi hanno avuto
una notevole importanza storica, ma al momento non sono pi\`u supportati.
Tra questi vogliamo ricordare i vari sistemi del progetto Automath \cite{Automath} di
De Bruijn, e il proof assistant Lego di Luo e Pollack \cite{lego92}.

\section{Alcuni risultati importanti}
Concludiamo questo articolo con una rassegna di alcuni importanti risultati
che, per varie ragioni, hanno costituito delle importanti pietre miliari
per la verifica automatica, o per l'evoluzione del particolare sistema di
prova. I risultati menzionati
sono spesso solamente i primi di una lunga serie di ulteriori approfondimenti;
lo stesso risultato \`e stato frequentemente ridimostrato con altri strumenti,
o con tecniche differenti. La lista \`e necessariamente incompleta: ci
scusiamo in anticipo con tutti gli autori i cui contributi, per ovvie ragioni
di spazio, non sono stati citati.

Suddividiamo i risultati in tre grossi gruppi: fondamenti (sia di matematica che
di informatica), teoremi di matematica avanzata, e applicazioni alla verifica
di hardware e software. 


\begin{center}
\begin{table}[htb]
\includegraphics[width=\textwidth]{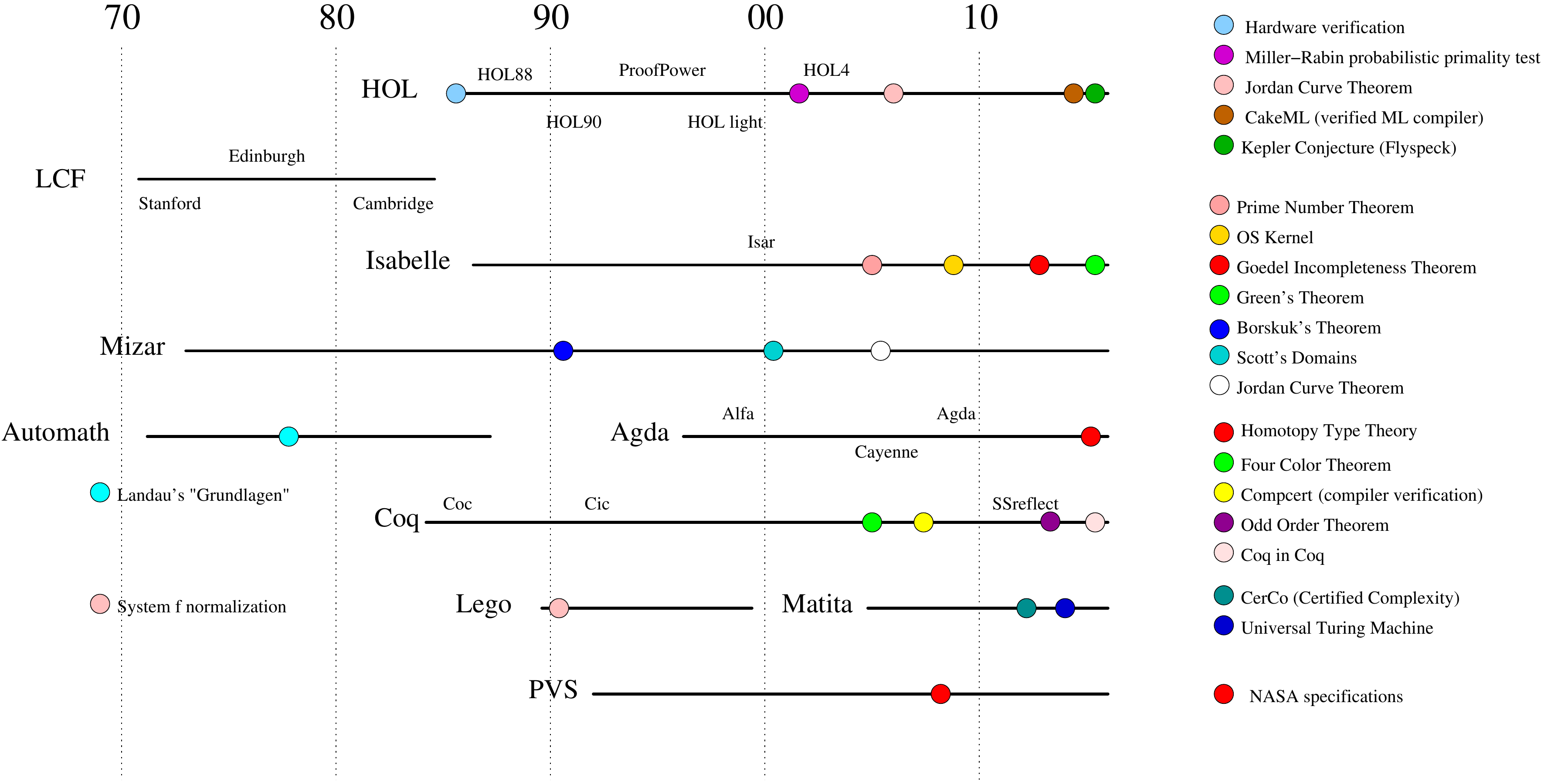}
\caption{Alcuni risultati notevoli}
\end{table}
\end{center}\vspace{-.5cm}

\subsection{Fondamenti di Matematica e Informatica}
Nel 1977, nella sua tesi di dottorato \cite{grundlagen}
sotto la direzione di De Bruijn,
van Benthem Jutting present\`o la completa formalizzazione e verifica
automatica dei {\em Grundlagen der Matematik} di Landau. Il risultato era
notevole e pass\`o per parecchio tempo insuperato, anche perch\`e LCF
stava cedendo il passo alla seconda generazione di dimostratori
interattivi: HOL,
Isabelle e Coq. Nella seconda met\`a degli anni ottanta e negli anni
novanta, sul fronte della formalizzazione, gran parte dei sistemi sono
impegnati a sviluppare le librerie matematiche di base sulla falsariga
dei grundlagen, ma andando ben oltre di questi. 

Un risultato notevole
dei primi anni novanta \`e la formalizzazione in Lego da parte di Berardi
della normalizzazione forte del sistema $F$ \cite{berardi1990girard}.
Il lavoro \`e importante sia perch\`e inaugura un filone fecondissimo
di studio formale della metateoria del lambda calcolo e dei meccanismi
di binding (notazione di De Bruijn diretta e inversa, tecniche nominali,
etc.), sia perch\`e rappresenta uno dei primi tentativi di dimostrare
la consistenza (di parte) del sistema logico dello strumento all'interno
del sistema stesso, ricerca recentemente culminata nel
monumentale lavoro di Barras \cite{coqincoq} di verifica della consistenza
di Coq all'interno di Coq stesso (a meno di un livello di universi).

Sempre sui fondamenti dell'informatica, un'altro lavoro notevole \`e la
formalizzazione della teoria dei domini di Scott fatta all'interno del
sistema Mizar da parte di Bancerek, Rudniki e
altri \cite{bancerek2002compendium}.

Per quanto riguarda la teoria della calcolabilti\`a, esiste una discreta
quantit\`a di
risultati sul lambda calcolo puro. L'esistenza di macchine di Turing universali
\`e stata certificata formalmente per la prima volta nel sistema
Matita nel 2015\cite{utm}. La teoria degli automi, cosi come la teoria
dei linguaggi formali \`e discretamente supportata in molti sistemi.

Sul fronte logico, il risultato pi\`u notevole \`e probabilmente la
formalizzazione da parte di Paulson in Isabelle dei
teoremi di incompletezza di G\"odel \cite{PaulsonGoedel}.

Un programma di ricerca particolarmente interessante \`e il complesso
lavoro di revisitazione fondazionale della matematica basato sulla
interpretazione omologica della teoria dei tipi (Univalent Foundations)
recentemente promosso da Voevodsky.  In Homotopy Type Theory
(HoTT) \cite{hottbook}, l'uguaglianza proposizionale \`e interpretata
come omotopia, e l'isomorfismo
di tipi come equivalenza omologica. I teoremi e le dimostrazioni in HoTT
hanno in questo modo una diretta interpretazione omologica.
L'Homotopy Type Theory si
\`e interamente sviluppata {\em congiuntamente} alla sua implementazione
formale: i sistemi che attualmente supportano HoTT sono Agda e Coq.

\subsection{Matematica}
I primi risultati matematici di una certa complessit\`a che sono
stati completamente formalizzati e verificati automaticamente
risalgono ai primi anno del nuovo millennio. Quasi contemporaneamente
vengono presentate due dimostrazioni alternative del teorema della
curva di Jordan (ogni curva chiusa del piano non intrecciata lo divide
in due parti, una ``interna'' e l'altra ``esterna''). La prima
dimostrazione \`e di Kornilowicz \cite{kornilowicz2005jordan} in Mizar,
la seconda di Hales \cite{Hales-jordan} in HOL-light.
Hales \`e un noto matematico, famoso sia per la sua dimostrazione della
congettura di Keplero sul modo pi\`u compatto per disporre delle sfere
in uno spazio euclideo tridimensionale, sia per gli eventi relativi
alla pubblicazione di questo risultato. Il punto era che la sua
dimostrazione comportava un largo uso di calcoli per risolvere
dei sistemi di vincoli che venivano delegati al computer. Dopo tre
anni di lavoro i revisori degli Annals of Mathematics, a cui il lavoro
era stato sottoposto per pubblicazione, concludevano che, bench\`e
ritenessero valido il lavoro, non erano in grado di certificarne la
correttezza dato il largo uso di ``low-level components'' su cui non
erano in grado di esprimere un giudizio. Alla fine gli Annals of Mathematics
pubblicarono una versione breve della prova \cite{Hales-kepler}, scorporando
le parti di codice. Una versione rivista e completa dell'articolo venne
in seguito pubblicata su Discrete and Computational Geometry. 
In seguito a questa esperienza Hales inizi\`o a interessarsi ai dimostratori
automatici (\cite{Hales-jordan} era un esercizio in questo senso), e
avvi\`o un grosso progetto denominato Flyspeck
(see \cite{NAMS-Hales}), per verificare in modo automatico e formale la
correttezza della sua dimostrazione. Il progetto si \`e concluso con successo
nel 2014 (il lavoro \`e in corso di pubblicazione). 

Altri due risultati importanti vengono raggiunti in quegli anni.
Nel 2007 \cite{Avigad07},
Avigad e altri dimostrano il teorema sulla distribuzione asintitoca
dei numeri primi. Quasi contemporaneamente, Gonthier presenta la sua
dimostrazione del teorema dei quattro colori \cite{Gonthier07}.
Per quest'ultimo teorema Gonthier sviluppa una interessante estensione
del linguaggio Coq (SSreflect \cite{ssreflect}) particolarmente orientata
alla manipolazione di strutture finite. Al fine di mostrare le
potenzialit\`a di questo linguaggio Gonthier decide di affrontare
il teorema di Feit-Thomson (ogni gruppo finito di ordine dispari
\`e risolubile), noto semplicemente come odd-order theorem. La
dimostrazione matematica ``cartacea'' \`e {\em mostruosa}: occupa
da sola un intero libro di 255 pagine ed \`e tutt'altro che {\em self contained}.
Anche in questo caso la formalizzazione \`e stata conclusa con
successo dopo tre anni di lavoro \cite{GonthierAll13}. \`E utile
osservare che la dimostrazione formale ha una dimensione comparabile
con quella cartacea.

\subsection{Verifica di hardware e software}
Le prime applicazioni rilevanti della verifica automatica
all'informatica si hanno nel campo dell'hardware. I sistemi della classe HOL
nascono in effetti dall'osservazione di Gordon e altri riguardo alla naturalezza
di impiegare meccanismi di ordine superiore nella specifica di propriet\`a
di circuiti logici \cite{BoultonGGHHT92}. Per lungo tempo la verifica si
incentra per lo pi\`u sulla correttezza di particolari algoritmi, e la
verifica di propriet\`a di protocolli. Tra gli esempi pi\`u rilevanti,
ricordiamo il lavoro di Hurd \cite{Hurd03} sul test di pimalit\`a
probabilistico di Miller-Rabin, che costituisce la prima formalizzazione
significativa nel campo della programmazione probabilistica.
In parallelo si affronta la semantica dei linguaggi di programmazione,
con importanti applicazioni alla sicurezza dei sistemi informativi
\cite{Leroy02,krakatoa}.

Nel 2006, suscita un immediato scalpore il lavoro di Leroy
\cite{leroy06,Leroy09} in cui per la prima volta si affronta la
correttezza di un programma software ragionevolmente complesso:
un compilatore ottimizzante per un sottoinsieme quasi completo del
linguaggio C, detto Cminor, verso codice assembler per PowerPC.
Il lavoro da luogo ad un grosso progetto di ricerca (Compcert)
che approfondisce l'argomento lungo svariate direzioni.

La tematica viene inoltre ripresa ed estesa anche da altri gruppi
di ricerca: ricordiamo il progetto CerCo \cite{CerCo-fopara} dove si affrontano
anche problematiche di complessit\`a computazionale del codice oggetto,
o il progetto CakeML \cite{CakeML} che fornisce per la prima volta un
compilatore certificato per il linguaggo SML.

Avendo acquisito confidenza sulla possibilit\`a pratica di certificare
in modo completo la correttezza del software, la ricerca si sta ora
allargando a problematiche via via pi\`u complesse, come ad esempio
la verifica di nuclei di sistemi operativi \cite{Klein10} e
sistemi critici di varia natura.


\end{document}